\RequirePackage{fix-cm}
\documentclass[twocolumn]{svjour3}          
\smartqed  
\usepackage{graphicx,float}
\usepackage{latexsym}
\usepackage[version=4]{mhchem} 
\usepackage{xcolor,ulem,hyperref}
\usepackage{booktabs}
\usepackage{multirow}
\usepackage{tabularx}
\usepackage{afterpage}
\usepackage{balance}
\usepackage{lastpage}
\usepackage{adjustbox}
\usepackage{dcolumn}
\usepackage{lineno}
\everymath{\displaystyle}
\journalname{Journal of Superconductivity and Novel Magnetism}
\begin{document}

\title{Peculiarities of niobium-based superconducting alloys in the light of crystal chemistry: A brief survey}

\titlerunning{Niobium-based intermetallic superconducting alloys}        

\author{Taimo Priinits         \and
        Artjom Vargunin \and Aleksandr Liivand 
}


\institute{T. Priinits \at
             Institute~of~Physics,~University~of~Tartu,~50411~Tartu,~Estonia \\
              \email{taimo.priinits@ut.ee}           
           \and
           A. Vargunin \at
           Institute~of~Physics,~University~of~Tartu,~50411~Tartu,~Estonia 
           \email{artjom.vargunin@ut.ee}
           \and
           A. Liivand \at
           Institute~of~Physics,~University~of~Tartu,~50411~Tartu,~Estonia
           \email{aleksandr.liivand@ut.ee}
}

\date{Received: date / Accepted: date}

\maketitle

\begin{abstract}
In this survey, we focus on \ce{Nb}-based binary intermetallic compounds,~which have been widely used over the last 40 years to develop a range of superconducting applications, including non-standard engineering solutions in the design of large magnets. However, since the 1980s, when it became apparent that the mechanical and superior superconducting properties of ordered intermetallic alloys such as \ce{Nb_{3}Sn} were largely due to their unique structural features, much of the research interest in the science of superconducting intermetallic alloys has been redirected to the development of necessary engineering applications in high magnetic field technology. Accordingly, the important role of crystal chemistry in understanding the fundamental aspects of the material properties of the \ce{Nb3Sn} family of intermetallics has been little explored. In the paper, we try to fill this gap by investigating the relationships between composition, microstructure and properties, highlighting their relevance to technological applications. Our goal is to link aspects of crystal chemistry with materials application issues. We shed light on the atomic assembly mechanisms and processes in terms of changes in the chemical environment, lattice structure, crystallization pathway, and macroscale phase textures, which can help in interpreting and explaining the prospects and limitations of the superconducting properties of \ce{Nb3Sn}. In the context of past and present prospects and limitations we briefly overview most important technological applications and discuss the various interrelations between superconductivity and structural properties of \ce{Nb}-based \ce{A{-15}} intermetallic alloys. We argue that these interrelations can be used to find the \ce{Nb}-based superconductors with more superior properties and stronger technological usability. 

\end{abstract}

\section{Introduction}
\label{intro}
At present, superconductivity technology offers innovative and large-scale solutions for various physical research applications: magnetic resonance imaging (MRI), nuclear magnetic resonance (NMR), fusion energy, quantum computing, etc. On the other hand, the potential of conventional superconductivity can be utilized even more efficiently, opening up new avenues for scientific progress and breakthroughs, as well as providing directions for modern technologies to achieve zero-emission goals, develop advanced medical diagnostics and therapeutics, expand energy capacity through the construction of power plants and stations, and create technologies for space applications.

In particular, the development of modern engineering designs and innovations places new demands on both the superconducting properties of alloys, their size, forming process, usable configuration and operational stability, and the strength properties of these materials to operate successfully in strong pulsed magnetic fields. Such materials are the cornerstone for realizing advances in superconducting technologies and applications. Accordingly, addressing material science aspects such as exploring novel intermetallic design concepts, improving the functional properties of superconducting materials, and developing new technologies to enhance the growth of critical superconducting properties is a strategic task in condensed matter physics and chemistry. Utilization of these properties opens wide opportunities for the application of superconducting compounds in scientific research and development. In modern technology there is an increasing need to apply strong magnetic fields on a large scale: in research facilities for controlled thermonuclear reactions, in plasma physics, in magnetohydrodynamic devices, in powerful motors and generators, etc. In all these cases, the use of superconducting magnets, wires and coils is the extremely promising cost-effective solution.

Among the known traditional superconducting materials, high-purity niobium and its derivatives, such as intermetallic alloys with \ce{A{-15}} type crystal structure~\cite{Matthias1954,Johnson1974,MATTHIAS1957138,Scanlan2004,Godeke2006_1,Stewart2015,Xu2017}, are still in large demand in the field of applied superconductivity. This is due to the unique combination of their superior superconducting properties, such as critical temperature that can exceed $18$ K, high critical current density, wide range of ideal diamagnetism, and high-frequency critical fields, which are currently unattainable for other superconductors. Such intermetallic alloys, which represent a very attractive object of solid-state research, have been the subject of numerous studies for several decades and the research interest is still at an increasing level~\cite{Narlikar2014}. In this context, in order to meet the growing technological demands, fundamental science is facing a number of challenges to find appropriate ways to improve the mechanical and physical properties of compound superconductors with \ce{A{-15}} structure. Therefore, the main purpose of this survey article is both to briefly highlight some actual aspects of the accumulated information in this field and to analyze possible relationship between fundamental research results and their present and future use in the development of high-tech superconducting applications.

Following a viewpoint that several fundamental properties of superconductivity are intimately linked to the composition-structure-property relations, in the first part of given paper we will consider the properties that play an important role in the enabling of superconductivity. In particular, we will summarize the specific features of the crystal-chemical background underlying the main properties of \ce{Nb}-based superconducting \ce{A{-15}} intermetallics. In order not to be overwhelmed by the vast amount of useful information obtained during more than half a century of fundamental research, we will focus on the correlation of the crystal-chemical background with structural and superconducting properties. There are many review articles in the literature, both old and new, dealing with various aspects of low temperature superconductivity and its applications~\cite{Johnson1974,DewHughes1975,Fietz1979,Braginski1985,Braginski1987,Xu_2017,Wu2020101410,Mitchell2009113}. However, information on how the interplay between chemical composition and crystalline architectures may determine key factors that could be responsible for the superconducting properties of \ce{A{-15}}-type materials is somewhat lacking in these papers. Motivated by this, we will try to complement these papers by discussing the crystal chemistry of these materials with respect to the macroscopic quantum properties of superconductivity in terms of what is understood and what is not. The second part deals with the applied aspects. By presenting several different engineering implementations as examples of current activities in applied superconductivity, we show why and how low-temperature superconductivity can play a key role in applied science and technology. The last section is devoted to the interdisciplinary research landscape of superconducting science, addressing applications in such fields as engineering, quantum computing, medicine, etc. Having briefly outlined the new horizons opened by scientific advances in some areas of modern superconducting electronics, we thus show that \ce{A{-15}}-type materials will remain at the center of the future technological development of modern superconducting applications.

\section{On structural chemistry of A-15-type superconducting intermetallics}
From a materials science point of view, a modern material is usually a bulk combination of two or more different chemical elements, due to the resulting structure achieving new properties that are not present in its individual components. In this respect, intermetallics are a unique class of materials characterized by the interplay of a metallic ground state and long-range ordering with high crystallographic symmetry. They are formed by the combination of two or more different elements which, in compositional ranges close to stoichiometric ratios, occupy the specific atomic positions in the crystal lattice during crystallization. Thus, the study of the processes that are responsible for the electronic and phononic properties and the strength of the interactions between them, in terms of phase and structural transformations, can provide important insights into a deep understanding of the superconducting mechanism in \ce{Nb_{3}X} compounds. The subsections presented below provides a characterization of the crystalline architecture of the binary intermetallic system and its internal microstructures. The survey focuses on crystal chemical parameters such as composition, lattice structure, stability and structural transformations, crystallographic patterns, which may be crucial for understanding the superconducting properties of an intermetallic system. To best illustrate the role that niobium plays in the \ce{A{-15}} structure, we also include an overview of our recent results~\cite{Taimo_2024} that cover several key aspects of alloy crystal chemistry in terms of settings, interpretation, characterization, and modeling. The key aspects are summarized in subsections 2.2 -- 2.4. This paper~\cite{Taimo_2024} gives a complete description of the methods of crystal chemical modeling and DFT-based simulations of structure and properties.

\subsection{Crystallographic data, stoichiometry and structural orderings}
\begin{figure*}[!htb]
\centering
  \includegraphics[width=0.67\textwidth]{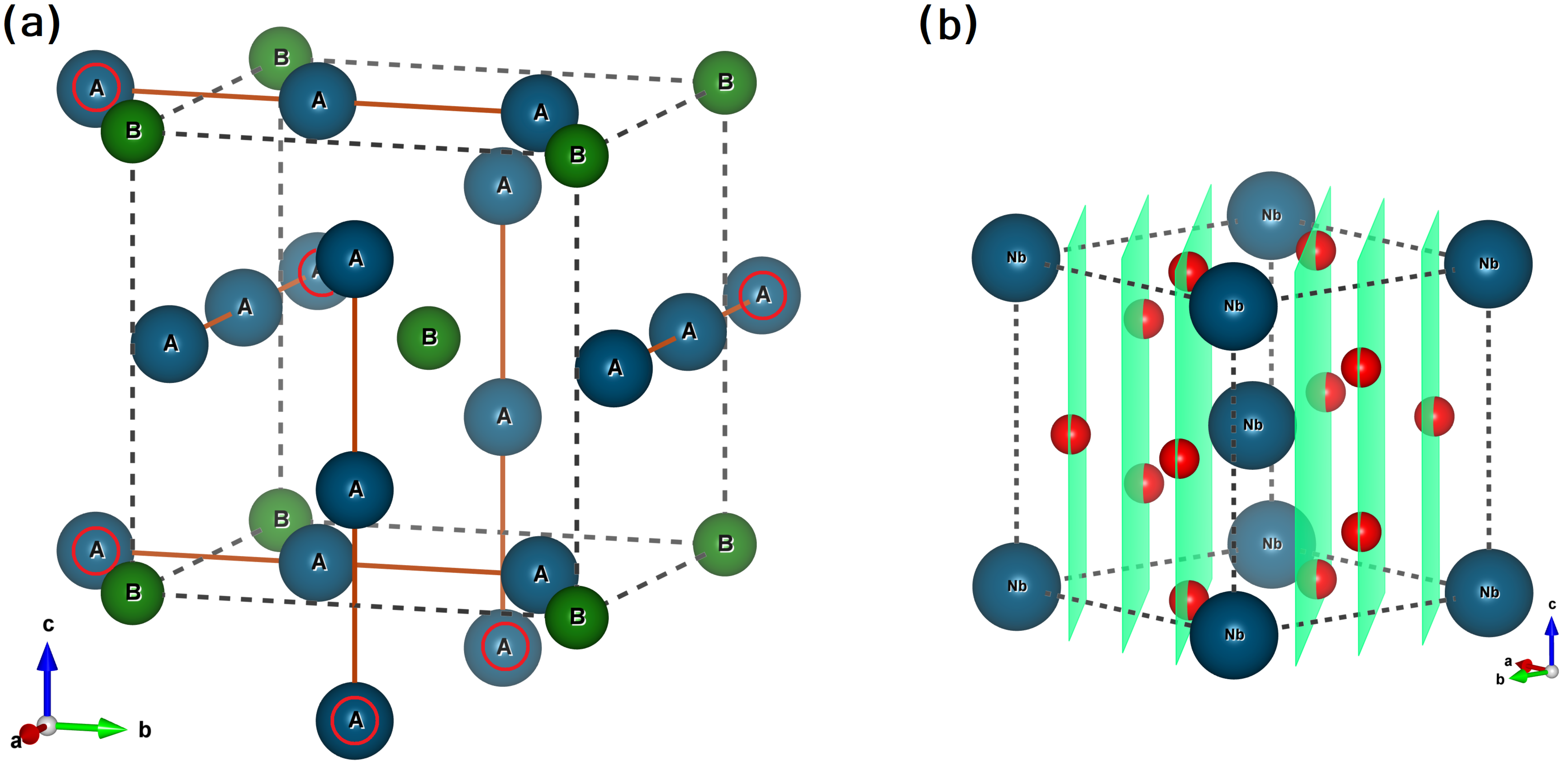}
\caption{
    Schematic comparison of structural atomic arrangements on metal and metalloid sublattices of the \ce{A3B} intermetallic compound. Left panel (a): The crystallographic illustration of the \ce{A{-15}} lattice structure, where the dashed gray lines outline the cubic unit cell, while the orange lines outline the sketch of mutually orthogonal chains of metal \ce{A} atoms. The neighboring cell \ce{A} atoms (additionally marked with red cycles) are added to the unit cell drawing to make the chains more transparent. The spatial $3D$ configuration of the \ce{B} atom sites reproduces the simple $bcc$ lattice.
    Right panel (b): Crystal-plane based reconstruction of the structural motif determined by the \ce{Nb} crystal lattice with symmetry of space group $Im3m$ (No. 229). In the unit cell shown, the model of active sites are represented in terms of the crystallographic interstitials that relate to the tetrahedral position with relative coordinates isomorphic to $(\frac14,0,\frac12)$ (shown as red balls). A set of crystal  planes isometric to the $(110)$ Miller plane is indicated in light green. The unit cells drawn in panels (a) and (b) are geometrically proportional.
}
\label{fig:unitcell}
\end{figure*}
The properties of intermetallic niobium alloys of binary composition have been the subject of extensive research during more than half a century; much useful information has become widely available~\cite{DewHughes1975,Fietz1979,Wu2020101410,Muller1980,Nevitt1967,Devantay1981}. As a background to our further work, in which niobium-containing compounds will be discussed in terms of the interplay between structural chemistry and properties, it may be instructive to first summarize some of their important crystal-chemical details. The \ce{A{-15}} crystal framework has a wide range of binary compositions with a nominal stoichiometric ratio of \ce{A_{3}B}. The \ce{A} atom, which is the base of the alloy, belongs to one of the transition metals such as \ce{Ti}, \ce{Nb}, \ce{V}, \ce{Zr}, \ce{Ta}, \ce{Cr} or \ce{Mo}. The \ce{B} atom represents a large number of alloying elements that include metals of groups IIIB and IVB, either one of the noble metals such as \ce{Pt}, \ce{Au}, \ce{Ge}, \ce{Os}, \ce{Ir}, etc., or one of the metalloids such as \ce{Al} or \ce{Sn}~\cite{Nevitt1967}. An interesting fact regarding the interactions that are responsible for a superconducting state is that the \ce{A{-15}} cubic structure is one of the close-packed crystal structures~\cite{DewHughes1975,Sun2023}. However, there must be a near stoichiometric balance between the \ce{A} and \ce{B} components, since violating such a specific compositional ratio on a macroscopic scale destabilizes the entire arrangement, making it completely impossible to form a stable \ce{A{-15}} phase~\cite{Muller1980,Rathz1981}.

The standard elemental description of the lattice structure and the composition of the stoichiometric composition of \ce{A3B} is given in the left panel (a) of Figure~\ref{fig:unitcell}. Upon crystallization, this composition acquires the \ce{A{-15}} cubic lattice, which corresponds to the \ce{Cr3Si} structure type~\cite{Nevitt1967}. As seen in Figure~\ref{fig:unitcell}(a), the peculiarity of the alloy lattice geometry is that each of the two constituents is accommodated on its own individual sublattice, whose crystallographic orbits are given by the $6c$ and $2a$ Wyckoff positions for the atoms \ce{A} and \ce{B}, respectively. In the framework of a more detailed crystallographic description, one can mention a characteristic feature of the lattice structure of \ce{A{-15}}, which is a quasi-one-dimensional arrangement of \ce{A} atoms~\cite{DewHughes1975,Muller1980,Glowacki1999}. In particular, \ce{A}-type atoms occupying positions with tetragonal site symmetry (point group $D_{2d}$) are uniformly arranged in pairs on the faces of the cubic lattice along three orthogonal axes. The atoms of each \ce{A} pair are separated by half the length of the unit cell; a total of six \ce{A} atoms are distributed at $6c$ Wyckoff position with a set of coordinates: {\small $(\frac14,0,\frac12)$, $(\frac34, 0,\frac12)$, $(\frac12,\frac14,0)$, $(\frac12,\frac34,0)$, $(0,\frac12,\frac14)$, and $(0,\frac12,\frac34)$}. On a larger scale, this arrangement results in atoms \ce{A} forming mutually orthogonal chains in directions parallel to the edges of unit cells, where the distance between these atoms in the chain is smallest compared to the distance between the nearest atoms belonging to different chains (as shown in Figure~\ref{fig:unitcell}(a) with the orange lines).

It has also been suggested that the short distance between nearest neighboring \ce{Nb} atoms in the \ce{A{-15}} crystal system could be responsible for generating some specific features of the electronic spectrum at the Fermi surface compared to the regular $bcc$ structure and the increase of the superconducting critical temperature~\cite{Godeke2006_1}, also atom size ratios and electronegativity are two of the most important parameters affecting the lattice constant~\cite{Muller1980}. To illustrate the effects from the side of \ce{A{-15}} lattice geometry, we have summarized some physical data in Table~\ref{tab:Nb3x_table}. This table shows the lattice constants and the charge per atom of some technologically important \ce{Nb_{3}X} compounds compared to \ce{Nb} and \ce{Nb-Ti}. It can be seen that the average number of valence electrons per atom is close to $5$ for the \ce{Nb_{3}X} systems. This would also support superconductivity according to the empirical Matthias rule~\cite{MATTHIAS1957138}.

\begin{table}[!htb]
\caption{Some physical properties of several technologically interesting intermetallics of the \ce{Nb_{3}X} family compared to the pure \ce{Nb} and \ce{NbTi} intermetallic alloy. The intermetallic behaviour is characterized by the value of lattice constant ($a$), resonance effect in terms of valence electron number per atom~\cite{MATTHIAS1957138}, superconducting temperature $T_c$, and critical magnetic field $H_{c2}$ at the temperature 4.2 K.}
\label{tab:Nb3x_table}
\begin{tabular}{l|llll}
\hline\noalign{\smallskip}
Compound & $a$ (\AA)              & $el/atom$ & $T_c$ (K)        & $H_{c2}$ (T)       \\
\noalign{\smallskip}\hline\noalign{\smallskip}
\ce{Nb3Al}      & $5.187$          & $4.5$     & $17.5$ \cite{Willens1969} & $31$ \cite{Stewart2015}   \\ [0.4ex]
\ce{Nb3Ga}      & $5.164$          & $4.5$     & $20.3$ \cite{DewHughes1975}  & $32$ \cite{Stewart2015}   \\ [0.4ex]
\ce{Nb3Ge}      & $5.166$          & $4.75$    & $23.0$ \cite{DewHughes1975}  & $36$ \cite{Fickett1985}   \\ [0.4ex]
\ce{Nb3In}      & $5.303$          & $4.5$     & $8$ \cite{DewHughes1975}     & $-$  \\ [0.4ex]
\ce{Nb3Sn}      & $5.289$          & $4.75$    & $18$ \cite{DewHughes1975}    & $23$ \cite{Fickett1985}   \\
\noalign{\smallskip}\hline\noalign{\smallskip}
\ce{Nb}      & 3.328 \cite{Schimmel_2005} & $6.5$     & $9.2$ \cite{Buschow2005}  & $0.16$ \cite{Finnemore1966} \\[0.4ex]
\ce{NbTi}    & 3.286 \cite{Baden_1983} & $4.5$     & $10$ \cite{Buschow2005}   & $11$ \cite{Fickett1985}  \\
\noalign{\smallskip}\hline
\end{tabular}
\end{table}
\subsection{Structural evolution and complexity}
The role of niobium, interpreted in terms of the complexity of the composition and the associated lattice site activities, has not yet been fully elucidated and this has prevented the completion of a consistent crystallographic picture. On the basis of a given general classification, it may be possible to explore in a qualitative way the specificity of the behavior of \ce{Nb} in the condensation of a binary alloy composition by using an appropriate structural matrix and then by modeling the corresponding solid phase relations. This problem has been considered in our work~\cite{Taimo_2024} under crystal chemical analysis using the relevant resources provided online by the Bilbao Crystallographic Server~\cite{bilbao_01,bilbao_02,bilbao_03,bilbao_04}. In particular, as illustrated in the right panel (b) of Figure \ref{fig:unitcell}, a set of special atomic positions can be introduced in order to compare the density of atomic packing between pure \ce{Nb} and its binary compound \ce{Nb_{3}Sn}. It should also be underlined that due to the specific features of the crystal architecture of niobium, the densest atomic packing in mixed binary solutions can be achieved directly by filling empty sites along the planes isometrically related to the generic plane $(110)$. Therefore, if the symmetry of the lattice allows certain crystallographic positions of atoms through which these planes can be drawn, then such atomic configurations, when assembled, will provide the necessary dense packing~\cite{Taimo_2024}. As shown in the (b) panel of Figure \ref{fig:unitcell}, in the cubic lattice of \ce{Nb}, the crystallographic voids of tetrahedral origin captured by $(110)$-type planes (marked as red balls in the image) can contribute to the formation of a tightly packed structure and thus offer the possible choice for an active site. 
\begin{table}[!htb]
\caption{Comparison of the compression properties of the pure \ce{Nb} metal and the cubic phase of the \ce{Nb_{3}Sn} intermetallic alloy shown in terms of elastic properties such as bulk modulus, $B$, and shear strength expressed by the difference of the stiffness tensor components, $(C_{11}-C_{12})/2$, (values are in GPa). $V_a$ denotes unit cell volume per atom. Numerical simulations of the structural and elastic behavior were performed using DFT calculations.}
\label{tab:Nb3Sn_elastic}
\abovetopsep=1.5pt
\centering
\resizebox{0.48\textwidth}{!}{ 
\begin{tabular}{@{}l|cccc@{}}
\hline\noalign{\smallskip}
System & $V_a$ & $B$ & $(C_{11}-C_{12})/2$ & $(C_{11}-C_{12})/3B$\\
\noalign{\smallskip}\hline\noalign{\smallskip}
\ce{Nb}  & $18.428$  & $170$ & $61.5$ & $0.24$\\[0.4ex]
\ce{Nb3Sn} (cubic) & $18.494$  & $166$ & $93.2$ & $0.37$ \\[0.4ex]
\noalign{\smallskip}\hline
\end{tabular}
}
\end{table}

The next question that has not been fully addressed in the literature is how the interaction between the alloy base and the incorporated alloying element, as well as the interaction between the alloying elements themselves, alters the metallic bonding that is an important factor in the stable constitution of the alloy base. Table~\ref{tab:Nb3Sn_elastic} attempts to shed some light on this question, at least on a qualitative level. The results of DFT simulations of the binary model system niobium-tin reveal several signs of atypical behavior. Obviously, these features are related to the internal structure of a material. In particular, the analysis of the elastic behavior before and after two-component intermetallic formation showed that the key parameters such as unit cell volume per atom and bulk moduli, which are affected by composition and lattice geometry factors, have found their way during transformations to restore their values to those determined by the prototype cubic structure of pure \ce{Nb}, which remained theirs as practically unchanged. Characteristics presented in the remaining two columns in terms of differences between diagonal and non-diagonal stiffness tensor components reflect the ductile/brittle classification~\cite{Senkov_2021}. That is, as for the comparison with \ce{Nb_{3}Sn} by means of such a recognition scheme, when the gap between $C_{11}$ and $C_{12}$ is widening, the ductility analysis indicates that the introduction of metalloid \ce{Sn} is accompanied by the formation of covalent (strongly directed) bonds. Accordingly, \ce{Nb_{3}Sn} is more brittle than solid \ce{Nb}. However, since directional bonding also contributes to this type of lattice ordering, this fact does not contradict the requirement of dense packing.
\subsection{On the crystallographic origins of \ce{Nb} chain formation}
\begin{figure*}[!htbp]
\centering
  \includegraphics[width=0.99\textwidth]{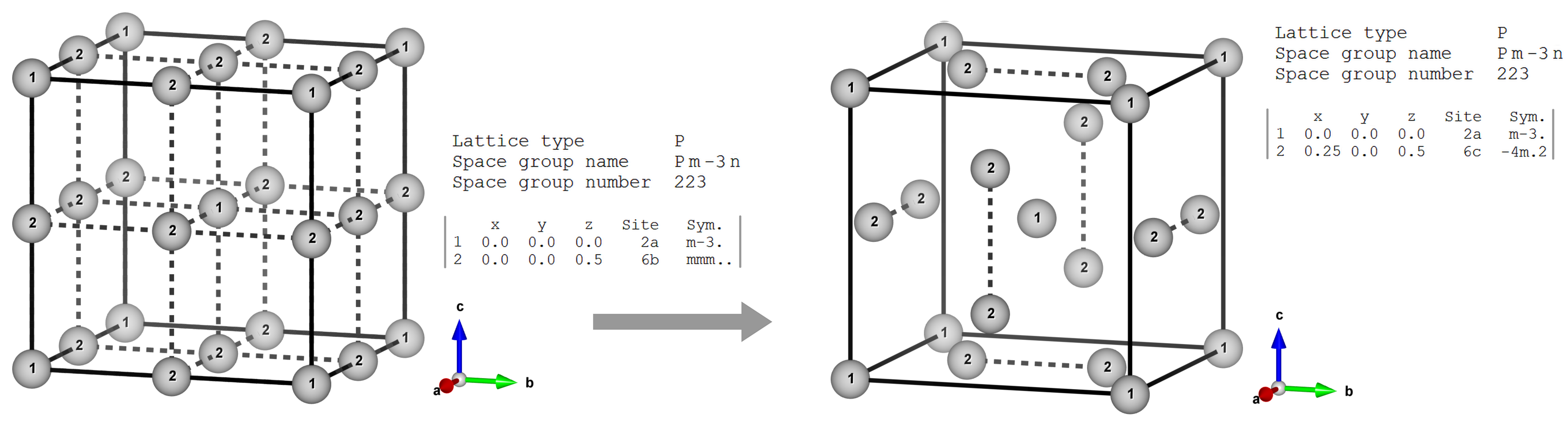}
\caption{Reconstructive model illustrating the formation of the \ce{A{-15}} lattice structure for the intermetallic compound \ce{Nb3Sn} and represented in terms of point symmetry changing ($D_{2h}$$\rightarrow$$D_{2d}$) initial positions of niobium atoms. The atomic-scale reconstruction scheme is drawn in terms of off-center displacements of niobium atoms along the three main crystallographic directions of the cubic architecture. The atomistic model was constructed and simulated using DFT calculations.}
\label{fig:evolution}
\end{figure*}
A comprehensive understanding of the processes of structural transformations within the strategy of rational modeling, which includes a close combination of both chemical principles and crystal geometry at the atomistic level, makes it possible to design realistic models on the basis of which it becomes capable to control how the evolution of electronic and phononic degrees of freedom occurs and to predict how the strength of their coupling changes. Obviously, this approach opens a wide way not only for describing many desirable properties of materials as a function of composition, structure, internal deformation, morphology, etc., but also opens new directions for describing the effects of physical forces, such as the mechanism of superconductivity. In the context of the present paper, this subsection gives a brief overview of our recent results on the structural chemistry of \ce{Nb3Sn} intermetallics~\cite{Taimo_2024}. In particular, based on the model of the ideal stoichiometric composition \ce{Nb3Sn}, we illustrate here the way in which the structural features of bulk metallic \ce{Nb} determine the distribution of \ce{Sn} and \ce{Nb} atoms in the \ce{A{-15}} lattice. Figure \ref{fig:evolution} represents a diagram sketch that gives a reconstruction of the mechanism of \ce{Nb} chain formation in terms of the partition configuration of \ce{Nb} atoms.

The left image of the sketch reflects the intermediate atomic configuration generated by the cubic distortion corresponding to the structural transformation $Im3m$ $\rightarrow$ $Pm$-$3n$. The main factors characterizing such a global change of the highly symmetric lattice structure are a fourfold increase of the unit cell volume and a decrease of the \ce{Nb} site symmetry in the distorted $Pm$-$3n$ lattice. However, due to crystal chemical effects, such a configuration is still poorly constrained because it remains less favorable in terms of stability. The right panel of figure \ref{fig:evolution} shows that the stability can be achieved by local displacements of niobium atoms. As a result, the identical chains of niobium atoms along three orthogonal cubic axes are generated. It is interesting to emphasize that such displacements partially restores the point symmetry of the \ce{Nb} sites from the input orthorhombic ($D_{2h}$) to the final tetragonal ($D_{2d}$). Thus, the existence of \ce{Nb} chains as a prominent feature of the \ce{Nb3Sn} system architecture is the direct consequence provided by the elemental \ce{Nb} bcc lattice.

\subsection{Induced anisotropy as an effect of the crystallographic orientation}
\begin{figure*}[!tp]
\centering
  \includegraphics[width=0.99\textwidth]{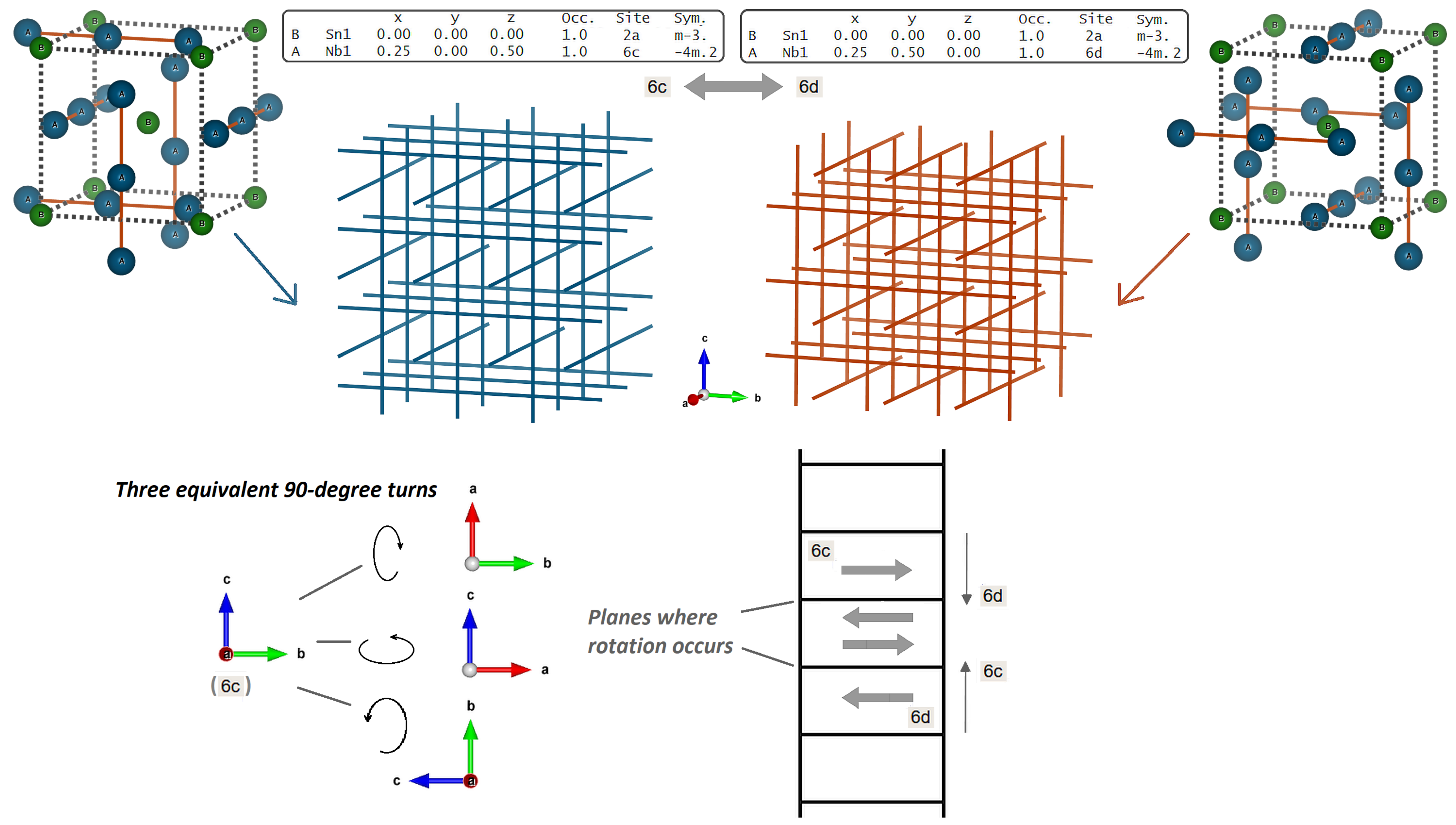}
\caption{Model of the bimodal internal microstructure associated with two lattice ordering motifs of \ce{Nb} chains predicted for the \ce{Nb3Sn} intermetallic compound. Atoms \ce{A} and \ce{B} of the \ce{A{-15}} cubic structure correspond to the usual distribution of \ce{Nb} and \ce{Sn} atoms. The illustrations of three-dimensional close-packing for both textures are shown schematically in terms of $6c$ (left) and $6d$ (right) Wyckoff positions in the unit cell of the \ce{A{-15}} cubic lattice, where the difference in crystallographic orientation of two texture patterns is determined by \ce{Nb} atomic chains formed by occupying the $6c$ (blue) and $6d$ (red) positions, respectively. At the bottom, the schematic arrangement of single crystal grains is shown for a polycrystalline system in terms of both textures. The grains, which differ in the orientation of the chains with respect to the three planes (as indicated in the lower left corner of the figure), are arbitrarily arranged along the main cubic crystallographic axes.}
\label{fig:comparison}
\end{figure*}
It has not been previously investigated whether the effect of crystallographic orientation contributes to the microstructural features introduced by the crystallization process. Crystal-chemical simulations predicted the effect of orientation-induced anisotropy for a \ce{Nb3Sn} polycrystalline intermetallic system~\cite{Taimo_2024}. An explanation is provided in Figure~\ref{fig:comparison}. To see how this is done, first note that the two crystallographic models are completely equivalent when it pertains to atomic-level characterization in the single crystalline regime. More precisely, in the case of single crystal crystallization, where the possibility of orientation distributions never arises, both the local structure and the lattice structure are accurately determined by one of these two models. On the other hand, for a polycrystalline intermetallic system, the role of the grain formation mechanism is well known. In terms of alloy solidification, this means that the local structure of each grain can be unambiguously determined from the crystallographic data shown in Figure~\ref{fig:comparison} (e.g. unit cell, point symmetries of occupied atomic positions, and bond geometry). However, the question arises as to how grains, which typically fill structured regions, can be classified at the microstructure scale in terms of their local lattice settings and atomic positions. The results of crystal chemical modeling have shown that in the equilibrium state, due to the requirement of equivariance with respect to the \ce{A{-15}} lattice architecture, two macroscopic textures that are not exactly equivalent will coexist in a polycrystalline material. Thus, orientation-induced anisotropy can be expected to occur at the crystallization stage when two or more crystalline grains merge. In this process, although their regular orientation with respect to the main lattice planes is preserved, a comparison between two separate grains from the microstructure composition indicates that they can be visually separated by rotating the atomic configuration by $90^{\circ}$ around the main crystallographic axis (by analogy, this is similar to making a half turn around the "twin" axis).
\begin{figure*}[!htb]
\centering
\includegraphics[width=0.50\textwidth]{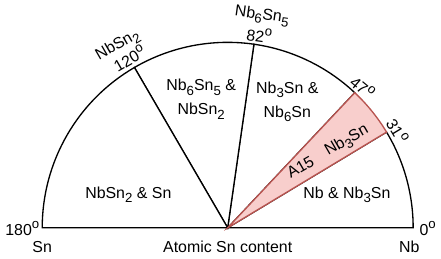}
\caption{Polar phase diagram schematically representing a series of intermediate compositions for the intermetallic system \ce{Nb{-}Sn} in terms of the ratio \ce{Nb}/\ce{Sn} changing as a polar angle. The compositional region where \ce{A{-15}} phase is stable is marked with red.
}
\label{fig:Nb3Sn_PhaseDiagram}
\end{figure*}
\subsection{Stoichiometric features}
Since the 70's, the niobium compounds of different stoichiometries have been extensively studied experimentally. The \ce{Nb} based \ce{A{-15}} compounds were found to be highly sensitive to these variations due to the stoichiometry changing electronic properties of original crystal system~\cite{Muller1980}. For example, \ce{Nb_{3}Ga}, $T_{c}$ varies from $14.5$ to $20.7$ K depending on the \ce{Nb}/\ce{Ga} ratio~\cite{Inoue2003}. For \ce{Nb_{3}Sn}, its $T_{c}$ varies from $6$ to $18$ K depending on the \ce{Sn} content~\cite{Posen2017}. The most common reason for off-stoichiometry effect is the irregular deposition of the substrate during \cite{Lee2018,Moore1979} fabrication. As mentioned above, the \ce{A{-15}} phase is technologically important, but it is only achievable within a certain range of stoichiometry. Figure \ref{fig:Nb3Sn_PhaseDiagram} illustrates this using the Nb-Sn intermetallic alloy as an example. \ce{Nb{-}Sn} becomes superconducting in the \ce{A{-15}} phase from about $18$ to $26$\% of Sn content \cite{Cooley2004,Godeke2006_1,Jo2014,Gala2016,Devantay1981,Moore1979}, with the highest critical temperature in the range of \ce{Nb}/\ce{Sn} $23$--$26$\%, which is therefore a practically usable region, because $ T_{c} $ starts to drop rapidly from $15$ K at $23$\% Sn content to $6$ K when shifting to 18\% Sn content \cite{Posen2017}.
More generally, it is the long-range order that is needed for a compound to have the best properties: if a material loses long-range order due to any kind of defect, its superconducting properties begin to degrade~\cite{Sun2023,Godeke2006_1,Posen2017,Gala2016,Besson2007,Li2017,Lee2020}. A certain amount of disorder has been observed to increase the lattice parameter~\cite{Stewart2015,Putti2008}, the electron mean free path~\cite{Sun2023}, the electron-phonon interaction strength~\cite{Godeke2006_2}, and the density of electron states near the Fermi level~\cite{Godeke2006_1,Gala2016,Putti2008}. It was also found that the lattice parameter is the main factor controlling the stoichiometric composition, and further reduction of the lattice parameter with external pressure leads to suppression of superconductivity~\cite{Stewart2015}.
\begin{figure*}[h!]
\centering
\includegraphics[width=0.95\textwidth]{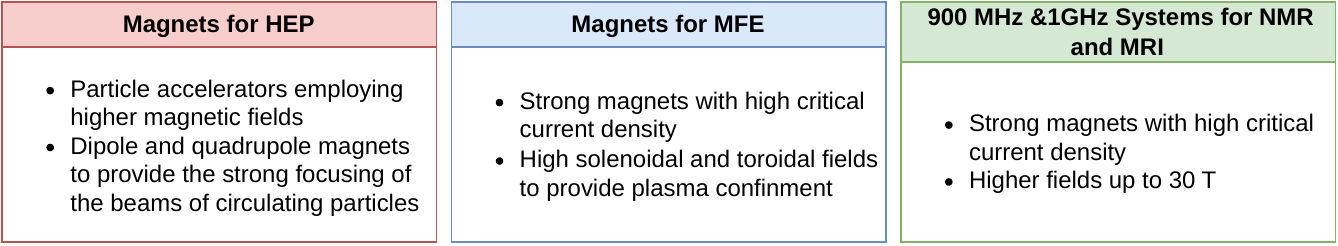}
\caption{Illustrative scheme showing the main directions of large-scale technological applications of \ce{Nb_{3}X} superconducting wires, tapes and cables. Abbreviations used: HEP -- high-energy physics, MFE -- magnetic fusion energy, the others are given in the text.}
\label{fig:applications_01v3}
\end{figure*}

\section{Applications}
The present section provides a brief overview of the applications of niobium alloys. Figure \ref{fig:applications_01v3} illustrates the main application directions for \ce{Nb3X} compounds. These directions are discussed in more detail below.
\subsection{{A{-15}} and other \ce{Nb} based superconductors}
\ce{A{-15}} compounds, namely \ce{Nb_{3}Sn}, \ce{Nb_{3}Ga} and \ce{Nb_{3}Al}, have been of interest for over 50 years for use in industrial applications due to their superior properties over pure \ce{Nb} and \ce{Nb} \ce{Ti} alloys, namely higher critical temperature, current density and critical magnetic field strength \cite{Fietz1979,Gavaler1975,Specking1993}. Table \ref{tab:Nb3x_table} summarizes some relevant properties of these systems. For large-scale applications, the production of \ce{Nb_{3}X} compounds with the required efficiency, i.e. production speed and quality, has been one of the major challenges to be overcome compared to regular \ce{Nb} and the easily processable and widely used \ce{Nb}-\ce{Ti} alloys. For example, \ce{Nb_{3}Sn} is a brittle material, and mechanical stress has a degrading effect on the superconducting properties, which makes it difficult to use \cite{Fietz1979,Specking1993,Glowacki1999}. Moreover, the superconducting coherence length of \ce{Nb_{3}Sn} is on the nanometer scale, what causes the properties to be highly dependent on even small defects~\cite{Sitaraman2023}.
\subsection{High Energy Physics}
\ce{Nb}-\ce{Ti} was the material of choice for older colliders, but the current density required for the next generation of colliders demands materials with better performance~\cite{Lee2003,Scanlan2004}. The current density for new colliders needs to be at least $ 10^{3} $ A/\ce{mm^{2}} in magnetic fields around $25$ T~\cite{Scanlan2004}. As planned, the design of the new Future Circular Collider will use dipole magnets with a magnetic field of $16$ T. Such a massive construction will require about 8000 tons of \ce{Nb_{3}Sn} wire, which also calls for the industry to scale up production \cite{Xu2017}. With the development of \ce{Nb_{3}Sn} wire for ITER, the scale-up for production has already begun.

Superconducting radiofrequency (SRF) cavities must be fabricated from materials with high critical temperature and large superconducting gap, low normal resistivity, $s$-wave gap symmetry, and high critical magnetic field~\cite{ValenteFeliciano2016}. New accelerators require magnets capable of operating with magnetic fields up to $11$ T \cite{Lee2003}. For $1.3$ GHz single-cell cavities, \ce{Nb_{3}Sn} has shown similar values at $4.2$ K compared to niobium at $2.0$ K, allowing a more energetically efficient design.     

\subsection{Fusion}
One of the largest fusion projects currently underway is the International Thermonuclear Experimental Reactor (ITER), where the tokamak magnet used in the design of ITER must generate magnetic fields of up to $12$ T and with a current density of $1100$ A/\ce{mm^{2}} at $4.2$ K~\cite{Flkiger2008}. The construction and development of ITER requires over 600 tons of \ce{Nb_{3}Sn}, the scale of this huge amount similar to the needs of the Future Circular Collider challenges the industry to greatly increase the manufacturing rates of \ce{Nb_{3}Sn} wire \cite{Zhou2011}.

\subsection{Scientific research, biology and medicine}
In contrast to industrial applications, \ce{Nb_{3}Sn} magnets have long been employed in scientific research facilities and equipment to test new theories and methods, approaches to plasma confinement, and to study the behavior of materials in high magnetic fields.
In biology and medicine, MRI and NMR are two of the most important methods relevant to a given topic and used for investigations. The design of next generation magnets (above $1$ GHz) is expected to use high critical currents with magnetic fields up to $30$ T \cite{Glowacki1999}. These parameters are not possible with normal \ce{Nb}-\ce{Ti} magnets and systems based on \ce{Nb_{3}Sn} or \ce{Nb_{3}Al} have to be used.

\subsection{Industry and economic benefits}
The benefits of using materials with higher current densities are that the physical dimensions of the facilities can be reduced, and the higher critical temperature makes it possible to operate superconducting devices at $4.2$ K instead of $2$ K.

Smaller facilities and higher operating temperatures reduce infrastructure, labor, and operation and maintenance costs, and also expand the range of industrial applications that use SRF accelerators (such as medical, border security, and flue gas and wastewater treatment). The ability to cool the material much farther below its critical temperature also increases the quality factor, improving its efficiency~\cite{Muzzi2015,Posen2021}. Given that the global superconductivity market is estimated to be in the billions of Eur, with the largest segment covering \ce{Nb}-\ce{Ti} based medical applications, the use of more advanced materials can therefore bring huge economic benefits.

\subsection{Future developments in applied superconductivity}
\begin{figure*}[!htp]
  \includegraphics[width=\textwidth]{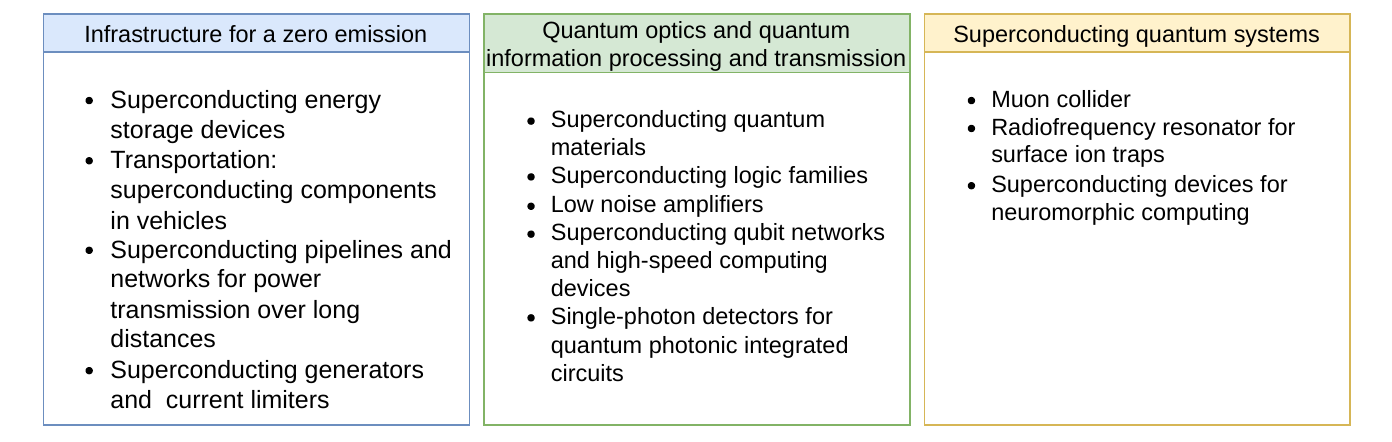}
\caption{Diagram showing new applications, outlook for future trends and new developments in superconducting technologies.}
\label{fig:applications_02v3}
\end{figure*}
Figure~\ref{fig:applications_02v3} shows a selection of applications that may become commercial as the properties of superconducting devices continue to advance, compared to the well-known, mostly large-scale applications described in the previous chapters. The first group of interest is the infrastructure for zero emission goals. For this, the superconducting energy storage devices are of interest due to the fact that these devices have much better efficiency, higher power density, charging speeds, etc., compared to the regular batteries or capacitors~\cite{Holla2015}. For superconducting components in vehicles, there is a direct interest to increase the capability of electric vehicles by reducing energy losses and increasing power, allowing to make smaller engines, with superconducting wires~\cite{Oyama2008}. One of the most talked about future applications is the use of superconducting wire in the power grid. In addition to efficiency, local communities would also appreciate the reduction of the electromagnetic field generated by power lines~\cite{Thomas2016}. Superconducting generators and current limiters improve the reliability and stability of the grid, helping to mitigate instabilities caused by faulty currents~\cite{Shirai2006}. The second group is concerned with quantum optics and information. Many quantum superconducting materials are reduced-dimensional materials in which electrons are confined in $2D$ layers. In these systems, electrons participate in the collective excitations that produce quantum effects \cite{Keimer2017}. Superconducting logic families use single flux quanta instead of voltage difference to transfer data. These logic gates can be optimized to operate at higher speeds or optimized to use less power for the same tasks (because at the current stage, most of the energy is dissipated as heat, and a fraction of the total energy is used to move information between logic gates)~\cite{Huang2022}. Low-noise amplifiers are devices mostly used for basic scientific measurements or for reading the qubit signal~\cite{Pagano2022}. Superconducting qubit networks and high-speed computing devices use the above logic gates to perform computations~\cite{Huang2022}. Single phonon detectors are devices used in quantum repeaters, qubit readers/amplifiers, quantum gates, and quantum computers to process quantum optical~information~\cite{Sprengers2011}. In the last group are devices for superconducting quantum systems. Muon colliders consist of precise measuring devices and high-energy accelerators to generate collisional muons; the lifetime of the muons is on the order of microseconds, making them difficult to study~\cite{Long2021}. Radiofrequency resonators for surface ion traps are devices that could be used to enable quantum computing and information processing with ions~\cite{Gandolfi2012}. The application of neural networks would be enhanced by superconducting devices for neuromorphic computing~\cite{Gandolfi2012}.
\section{Conclusion}
After the discovery of the phenomenon of superconductivity in 1911, nearly half a century passed before the possibilities of material design and engineering of functional superconducting systems for the development of a wide range of applications in science, industry, medicine, and quantum computing were realized. The development of such specific materials with high sensitivity to magnetic flux has been successful, as numerous attempts by researchers over several decades of basic research have yielded suitable superconducting compounds with significant advantages for many large-scale cryogenic applications. Two key materials that have been shown to be capable of stable operation in liquid helium environments have received much attention in these developments: (i) in high energy physics, \ce{NbTi} superconductors, which have shown technical success in accelerated particle physics, and (ii) \ce{Nb3Sn} superconductors, which are the established basis for the design of large high field magnets.

The family of \ce{A{-15}} intermetallic alloys includes \ce{A_{3}B} compositions possessing ideal stoichiometric ratios. In this family, compounds based on \ce{Nb} \ce{Nb_{3}X} have been of scientific and technological interest for more than 50 years. The \ce{Nb3Sn} system, together with other \ce{Nb}-based \ce{A{-15}} intermetallics, has been mainly considered in this brief survey. The main peculiarity of our work is that, in order to analyze a number of key properties of \ce{Nb} intermetallics, we have made a special projection, based on crystal chemistry, between the crystallography and the ground state properties of these binary compounds. As such an approach clarified aspects of our interest related to the interplay of chemistry and crystallography, we tried to summarize some information related to the current state of technological applications of \ce{A{-15}} superconducting systems. Based on the analysis and simulations performed, we have filled the gap in the understanding of how the spatial rearrangement of Nb atoms into chains occurs during the formation of the \ce{Nb}-based \ce{A{-15}} structure. Since these chains are considered by the scientific community to be conducive to the superconducting phenomenon in \ce{A{-15}} intermetallic alloys, we have briefly outlined the results of our modeling of the evolution of the \ce{A{-15}} lattice structure. We have shown that the formation of niobium chains, due to the crystallographic effect, is a necessary condition for the stabilization of the cubic lattice of \ce{Nb3Sn} after the transformation from the highly symmetric lattice prototype. The other crystallographic effect we discussed is related to the phenomenon of structural shift between orientations of niobium chains. We have argued that this is a specific feature of the \ce{A{-15}} lattice caused by the $6c\,{\rightarrow}\,6d$ bias in the occupation of the $6c$ Wyckoff positions. This effect occurs only at the microstructure level, where the orientation of the niobium chains leads to anisotropy due to the distinct chain patterning in the neighboring crystal grains of the polycrystalline \ce{A{-15}} system.

We have also surveyed how composition, stoichiometric content and lattice properties can be connected with superconducting properties. Briefly, we started with a general description of the crystal structure, namely how the atoms are arranged in the lattice unit cell (Figure \ref{fig:unitcell}). This was followed by an overview of how the formation of \ce{Nb3Sn} can be represented from a crystallographic point of view within crystal structure theory. The basic modeling was used to observe several important crystal chemical aspects, such as the atomic behavior during the lattice transformations. Then, some general peculiarities of the \ce{A{-15}} structure were summarized - how the nearest neighbor distance might favor superconductivity by inducing peculiarities in the electronic spectrum of the Fermi level. It has also been highlighted that a deep understanding of the relationships between compositional, structural, and superconducting properties could provide insight into how to design superconductors with desirable superior properties and that belong to the most technologically exploited family of materials.
\begin{acknowledgements}
\end{acknowledgements}

%
\section*{Conflict of interest}
The authors declare that they have no conflict of interest.

\bibliographystyle{spphys}       
\bibliography{paper}   

\begin{thebibliography}{10}
\providecommand{\url}[1]{{#1}}
\providecommand{\urlprefix}{URL }
\expandafter\ifx\csname urlstyle\endcsname\relax
  \providecommand{\doi}[1]{DOI \discretionary{}{}{}#1}\else
  \providecommand{\doi}{DOI \discretionary{}{}{}\begingroup
  \urlstyle{rm}\Url}\fi

\bibitem{Matthias1954}
B.T. Matthias, T.H. Geballe, S.~Geller, E.~Corenzwit, Phys. Rev. \textbf{95},
  1435 (1954)

\bibitem{Johnson1974}
G.R. Johnson, D.H. Douglass, J. of Low Temp. Phys. \textbf{14}, 565 (1974)

\bibitem{MATTHIAS1957138}
B.~Matthias,  (Elsevier, 1957), pp. 138--150.
\newblock \doi{https://doi.org/10.1016/S0079-6417(08)60104-3}.
\newblock
  \urlprefix\url{https://www.sciencedirect.com/science/article/pii/S0079641708601043}

\bibitem{Scanlan2004}
R.~Scanlan, A.~Malozemoff, D.~Larbalestier, Proceedings of the IEEE
  \textbf{92}(10), 1639 (2004).
\newblock \doi{10.1109/jproc.2004.833673}.
\newblock \urlprefix\url{http://dx.doi.org/10.1109/JPROC.2004.833673}

\bibitem{Godeke2006_1}
A.~Godeke, Superconductor Science and Technology \textbf{19}(8), R68 (2006).
\newblock \doi{10.1088/0953-2048/19/8/r02}.
\newblock \urlprefix\url{http://dx.doi.org/10.1088/0953-2048/19/8/R02}

\bibitem{Stewart2015}
G.~Stewart, Physica C: Superconductivity and its Applications \textbf{514}, 28
  (2015).
\newblock \doi{10.1016/j.physc.2015.02.013}.
\newblock \urlprefix\url{http://dx.doi.org/10.1016/j.physc.2015.02.013}

\bibitem{Xu2017}
X.~Xu, Superconductor Science and Technology \textbf{30}(9), 093001 (2017).
\newblock \doi{10.1088/1361-6668/aa7976}.
\newblock \urlprefix\url{http://dx.doi.org/10.1088/1361-6668/aa7976}

\bibitem{Narlikar2014}
A.V. Narlikar, \emph{Superconductors 1st edn} (Oxford University Press, 2014)

\bibitem{DewHughes1975}
D.~Dew-Hughes, Cryogenics \textbf{15}(8), 435 (1975).
\newblock \doi{10.1016/0011-2275(75)90019-3}.
\newblock \urlprefix\url{http://dx.doi.org/10.1016/0011-2275(75)90019-3}

\bibitem{Fietz1979}
W.~Fietz, IEEE Transactions on Magnetics \textbf{15}(1), 67 (1979).
\newblock \doi{10.1109/tmag.1979.1060147}.
\newblock \urlprefix\url{http://dx.doi.org/10.1109/TMAG.1979.1060147}

\bibitem{Braginski1985}
A.I. Braginski, J.R. Gavaler, \emph{Superconducting Electronic Film Structures,
  Semiannual Report} (Westinghouse R{\&}D Center, 1985)

\bibitem{Braginski1987}
A.I. Braginski, J.R. Gavaler, \emph{Superconducting Electronic Film Structures,
  Annual Report} (Westinghouse R{\&}D Center, 1987)

\bibitem{Xu_2017}
X.~Xu, Superconductor Science and Technology \textbf{30}(9), 093001 (2017).
\newblock \doi{10.1088/1361-6668/aa7976}.
\newblock \urlprefix\url{https://dx.doi.org/10.1088/1361-6668/aa7976}

\bibitem{Wu2020101410}
Y.~Wu, L.~Bao, X.~Wang, Y.~Wang, M.~Peng, Y.~Duan, Materials Today
  Communications \textbf{25}, 101410 (2020).
\newblock \doi{https://doi.org/10.1016/j.mtcomm.2020.101410}.
\newblock
  \urlprefix\url{https://www.sciencedirect.com/science/article/pii/S2352492820324211}

\bibitem{Mitchell2009113}
N.~Mitchell, P.~Bauer, D.~Bessette, A.~Devred, R.~Gallix, C.~Jong, J.~Knaster,
  P.~Libeyre, B.~Lim, A.~Sahu, F.~Simon, Fusion Engineering and Design
  \textbf{84}(2), 113 (2009).
\newblock \doi{https://doi.org/10.1016/j.fusengdes.2009.01.006}.
\newblock
  \urlprefix\url{https://www.sciencedirect.com/science/article/pii/S0920379609000301}.
\newblock Proceeding of the 25th Symposium on Fusion Technology

\bibitem{Taimo_2024}
T.~Priinits, A.~Vargunin, A.~Liivand,  (2024).
\newblock To be published

\bibitem{Muller1980}
J.~Muller, Reports on Progress in Physics \textbf{43}(5), 641 (1980).
\newblock \doi{10.1088/0034-4885/43/5/003}.
\newblock \urlprefix\url{http://dx.doi.org/10.1088/0034-4885/43/5/003}

\bibitem{Nevitt1967}
N.~Nevitt, in \emph{Intermetallic compounds}, ed. by J.H. Westbrook, Wiley
  Series on the Science and Technology of Materials (John Willey and Sons,
  1967)

\bibitem{Devantay1981}
H.~Devantay, J.L. Jorda, M.~Decroux, J.~Muller, R.~Flükiger, Journal of
  Materials Science \textbf{16}(8), 2145 (1981).
\newblock \doi{10.1007/bf00542375}.
\newblock \urlprefix\url{http://dx.doi.org/10.1007/BF00542375}

\bibitem{Sun2023}
Z.~Sun, Z.~Baraissov, R.D. Porter, L.~Shpani, Y.T. Shao, T.~Oseroff, M.O.
  Thompson, D.A. Muller, M.U. Liepe, Superconductor Science and Technology
  \textbf{36}(11), 115003 (2023).
\newblock \doi{10.1088/1361-6668/acf5ab}.
\newblock \urlprefix\url{http://dx.doi.org/10.1088/1361-6668/acf5ab}

\bibitem{Rathz1981}
T.J. {Rathz}.
\newblock Susceptibility measurements on the superconducting properties of
  nb-ge alloys (1981).
\newblock
  \urlprefix\url{https://ui.adsabs.harvard.edu/abs/1981smsp.rept.....R}.
\newblock Provided by the SAO/NASA Astrophysics Data System

\bibitem{Glowacki1999}
B.~Glowacki, Intermetallics \textbf{7}(2), 117 (1999).
\newblock \doi{10.1016/s0966-9795(98)00084-3}.
\newblock \urlprefix\url{http://dx.doi.org/10.1016/S0966-9795(98)00084-3}

\bibitem{Willens1969}
R.~Willens, T.~Geballe, A.~Gossard, J.~Maita, A.~Menth, G.~Hull, R.~Soden,
  Solid State Communications \textbf{7}(11), 837 (1969).
\newblock \doi{10.1016/0038-1098(69)90773-x}.
\newblock \urlprefix\url{http://dx.doi.org/10.1016/0038-1098(69)90773-X}

\bibitem{Fickett1985}
F.~Fickett, Journal of Research of the National Bureau of Standards
  \textbf{90}(2), 95 (1985).
\newblock \doi{10.6028/jres.090.007}.
\newblock \urlprefix\url{http://dx.doi.org/10.6028/jres.090.007}

\bibitem{Schimmel_2005}
H.G. Schimmel, J.~Huot, L.C. Chapon, F.D. Tichelaar, F.M. Mulder, Journal of
  the American Chemical Society \textbf{127}(41), 14348–14354 (2005).
\newblock \doi{10.1021/ja051508a}.
\newblock \urlprefix\url{http://dx.doi.org/10.1021/ja051508a}

\bibitem{Buschow2005}
K.~Buschow, \emph{Concise Encyclopedia of Magnetic and Superconducting
  Materials}.
\newblock Advances in Materials Sciences and Engineering (Elsevier Science,
  2005).
\newblock \urlprefix\url{https://books.google.ee/books?id=N9mvytGEBtwC}

\bibitem{Finnemore1966}
D.K. Finnemore, T.F. Stromberg, C.A. Swenson, Phys. Rev. \textbf{149}, 231
  (1966).
\newblock \doi{10.1103/PhysRev.149.231}.
\newblock \urlprefix\url{https://link.aps.org/doi/10.1103/PhysRev.149.231}

\bibitem{Baden_1983}
W.~Baden, A.~Weiss, International Journal of Materials Research \textbf{74}(2),
  89–93 (1983).
\newblock \doi{10.1515/ijmr-1983-740207}.
\newblock \urlprefix\url{http://dx.doi.org/10.1515/ijmr-1983-740207}

\bibitem{bilbao_01}
M.~Aroyo, J.~P{\'{e}}rez-Mato, D.~Orobengoa, E.~Tasci, G.~de~la Flor, A.~Kirov,
  Bulg. Chem. Commun. \textbf{43}(2), 183 (2011)

\bibitem{bilbao_02}
M.~Aroyo, J.~P{\'{e}}rez-Mato, C.~Capillas, E.~Kroumova, S.~Ivantchev,
  G.~Madariaga, A.~Kirov, H.~Wondratschek, Z. Krist. \textbf{221}(1), 15 (2006)

\bibitem{bilbao_03}
M.I. Aroyo, A.~Kirov, C.~Capillas, J.M. P{\'{e}}rez-Mato, H.~Wondratschek, Acta
  Cryst. \textbf{A62}(2), 115 (2006).
\newblock \doi{10.1107/S0108767305040286}

\bibitem{bilbao_04}
S.~Ivantchev, E.~Kroumova, G.~Madariaga, J.M. P{\'{e}}rez-Mato, M.I. Aroyo, J.
  Appl. Crystallogr. \textbf{33}(4), 1190 (2000).
\newblock \doi{10.1107/S0021889800007135}

\bibitem{Senkov_2021}
O.N. Senkov, D.B. Miracle, Scientific Reports \textbf{11}(1) (2021).
\newblock \doi{10.1038/s41598-021-83953-z}.
\newblock \urlprefix\url{http://dx.doi.org/10.1038/s41598-021-83953-z}

\bibitem{Inoue2003}
K.~Inoue, A.~Kikuchi, Y.~Yoshida, Y.~Iijima, Physica C: Superconductivity
  \textbf{384}(3), 267–273 (2003).
\newblock \doi{10.1016/s0921-4534(02)01879-8}.
\newblock \urlprefix\url{http://dx.doi.org/10.1016/S0921-4534(02)01879-8}

\bibitem{Posen2017}
S.~Posen, D.L. Hall, Superconductor Science and Technology \textbf{30}(3),
  033004 (2017).
\newblock \doi{10.1088/1361-6668/30/3/033004}.
\newblock \urlprefix\url{http://dx.doi.org/10.1088/1361-6668/30/3/033004}

\bibitem{Lee2018}
J.~Lee, S.~Posen, Z.~Mao, Y.~Trenikhina, K.~He, D.L. Hall, M.~Liepe, D.N.
  Seidman, Superconductor Science and Technology \textbf{32}(2), 024001 (2018).
\newblock \doi{10.1088/1361-6668/aaf268}.
\newblock \urlprefix\url{http://dx.doi.org/10.1088/1361-6668/aaf268}

\bibitem{Moore1979}
D.F. Moore, R.B. Zubeck, J.M. Rowell, M.R. Beasley, Phys. Rev. B \textbf{20},
  2721 (1979).
\newblock \doi{10.1103/PhysRevB.20.2721}.
\newblock \urlprefix\url{https://link.aps.org/doi/10.1103/PhysRevB.20.2721}

\bibitem{Cooley2004}
L.D. Cooley, C.M. Fischer, P.J. Lee, D.C. Larbalestier, Journal of Applied
  Physics \textbf{96}(4), 2122 (2004).
\newblock \doi{10.1063/1.1763993}.
\newblock \urlprefix\url{http://dx.doi.org/10.1063/1.1763993}

\bibitem{Jo2014}
Y.J. Jo, J.~Zhou, Z.H. Sung, P.J. Lee, D.C. Larbalestier, APL Materials
  \textbf{2}(10) (2014).
\newblock \doi{10.1063/1.4896935}.
\newblock \urlprefix\url{http://dx.doi.org/10.1063/1.4896935}

\bibitem{Gala2016}
F.~Gala, G.~De~Marzi, L.~Muzzi, G.~Zollo, Physical Chemistry Chemical Physics
  \textbf{18}(48), 32840 (2016).
\newblock \doi{10.1039/c6cp06699b}.
\newblock \urlprefix\url{http://dx.doi.org/10.1039/c6cp06699b}

\bibitem{Besson2007}
R.~Besson, S.~Guyot, A.~Legris, Phys. Rev. B \textbf{75}, 054105 (2007).
\newblock \doi{10.1103/PhysRevB.75.054105}.
\newblock \urlprefix\url{https://link.aps.org/doi/10.1103/PhysRevB.75.054105}

\bibitem{Li2017}
Y.~Li, Y.~Gao, Scientific Reports \textbf{7}(1) (2017).
\newblock \doi{10.1038/s41598-017-01292-4}.
\newblock \urlprefix\url{http://dx.doi.org/10.1038/s41598-017-01292-4}

\bibitem{Lee2020}
J.~Lee, Z.~Mao, K.~He, Z.H. Sung, T.~Spina, S.I. Baik, D.L. Hall, M.~Liepe,
  D.N. Seidman, S.~Posen, Acta Materialia \textbf{188}, 155 (2020).
\newblock \doi{10.1016/j.actamat.2020.01.055}.
\newblock \urlprefix\url{http://dx.doi.org/10.1016/j.actamat.2020.01.055}

\bibitem{Putti2008}
M.~Putti, R.~Vaglio, J.M. Rowell, Superconductor Science and Technology
  \textbf{21}(4), 043001 (2008).
\newblock \doi{10.1088/0953-2048/21/4/043001}.
\newblock \urlprefix\url{http://dx.doi.org/10.1088/0953-2048/21/4/043001}

\bibitem{Godeke2006_2}
A.~Godeke, B.t. Haken, H.H.J.t. Kate, D.C. Larbalestier, Superconductor Science
  and Technology \textbf{19}(10), R100 (2006).
\newblock \doi{10.1088/0953-2048/19/10/r02}.
\newblock \urlprefix\url{http://dx.doi.org/10.1088/0953-2048/19/10/R02}

\bibitem{Gavaler1975}
J.~Gavaler, M.~Janocko, A.~Braginski, G.~Roland, IEEE Transactions on Magnetics
  \textbf{11}(2), 192 (1975).
\newblock \doi{10.1109/tmag.1975.1058651}.
\newblock \urlprefix\url{http://dx.doi.org/10.1109/TMAG.1975.1058651}

\bibitem{Specking1993}
W.~Specking, H.~Kiesel, H.~Nakajima, T.~Ando, H.~Tsuji, Y.~Yamada, M.~Nagata,
  IEEE Transactions on Applied Superconductivity \textbf{3}(1), 1342 (1993).
\newblock \doi{10.1109/77.233651}.
\newblock \urlprefix\url{http://dx.doi.org/10.1109/77.233651}

\bibitem{Sitaraman2023}
N.S. Sitaraman, Z.~Sun, B.L. Francis, A.C. Hire, T.~Oseroff, Z.~Baraissov, T.A.
  Arias, R.G. Hennig, M.U. Liepe, D.A. Muller, M.K. Transtrum, Phys. Rev. Appl.
  \textbf{20}, 014064 (2023).
\newblock \doi{10.1103/PhysRevApplied.20.014064}.
\newblock
  \urlprefix\url{https://link.aps.org/doi/10.1103/PhysRevApplied.20.014064}

\bibitem{Lee2003}
P.~Lee, D.~Larbalestier, in \emph{Proceedings of the 2003 Bipolar/BiCMOS
  Circuits and Technology Meeting (IEEE Cat. No.03CH37440)}, \emph{PAC-03},
  vol.~1 (IEEE, 2003), \emph{PAC-03}, vol.~1, pp. 151 -- 155 Vol.1.
\newblock \doi{10.1109/pac.2003.1288865}.
\newblock \urlprefix\url{http://dx.doi.org/10.1109/PAC.2003.1288865}

\bibitem{ValenteFeliciano2016}
A.M. Valente-Feliciano, Superconductor Science and Technology \textbf{29}(11),
  113002 (2016).
\newblock \doi{10.1088/0953-2048/29/11/113002}.
\newblock \urlprefix\url{http://dx.doi.org/10.1088/0953-2048/29/11/113002}

\bibitem{Flkiger2008}
R.~Fl\"{u}kiger, D.~Uglietti, C.~Senatore, F.~Buta, Cryogenics
  \textbf{48}(7-8), 293 (2008).
\newblock \doi{10.1016/j.cryogenics.2008.05.005}.
\newblock \urlprefix\url{http://dx.doi.org/10.1016/j.cryogenics.2008.05.005}

\bibitem{Zhou2011}
J.~Zhou, Y.~Jo, Z.~Hawn~Sung, H.~Zhou, P.J. Lee, D.C. Larbalestier, Applied
  Physics Letters \textbf{99}(12) (2011).
\newblock \doi{10.1063/1.3643055}.
\newblock \urlprefix\url{http://dx.doi.org/10.1063/1.3643055}

\bibitem{Muzzi2015}
L.~Muzzi, G.~De~Marzi, A.~Di~Zenobio, A.~della Corte, Superconductor Science
  and Technology \textbf{28}(5), 053001 (2015).
\newblock \doi{10.1088/0953-2048/28/5/053001}.
\newblock \urlprefix\url{http://dx.doi.org/10.1088/0953-2048/28/5/053001}

\bibitem{Posen2021}
S.~Posen, J.~Lee, D.N. Seidman, A.~Romanenko, B.~Tennis, O.S. Melnychuk, D.A.
  Sergatskov, Superconductor Science and Technology \textbf{34}(2), 025007
  (2021).
\newblock \doi{10.1088/1361-6668/abc7f7}.
\newblock \urlprefix\url{http://dx.doi.org/10.1088/1361-6668/abc7f7}

\bibitem{Holla2015}
R.V. Holla, Journal of Undergraduate Research \textbf{5}(1), 49 (2015)

\bibitem{Oyama2008}
H.~Oyama, T.~Shinzato, K.~Hayashi, K.~Kitajima, T.~Ariyoshi, T.~Sawai, SEI
  Tech. Rev \textbf{67}, 22 (2008)

\bibitem{Thomas2016}
H.~Thomas, A.~Marian, A.~Chervyakov, S.~St\"{u}ckrad, D.~Salmieri, C.~Rubbia,
  Renewable and Sustainable Energy Reviews \textbf{55}, 59 (2016).
\newblock \doi{10.1016/j.rser.2015.10.041}.
\newblock \urlprefix\url{http://dx.doi.org/10.1016/j.rser.2015.10.041}

\bibitem{Shirai2006}
Y.~Shirai, T.~Morimoto, K.~Furushiba, M.~Shiotsu, K.~Fushiki, J.~Baba,
  T.~Nitta, Journal of Physics: Conference Series \textbf{43}, 958–961
  (2006).
\newblock \doi{10.1088/1742-6596/43/1/234}.
\newblock \urlprefix\url{http://dx.doi.org/10.1088/1742-6596/43/1/234}

\bibitem{Keimer2017}
B.~Keimer, J.E. Moore, Nature Physics \textbf{13}(11), 1045 (2017).
\newblock \doi{10.1038/nphys4302}.
\newblock \urlprefix\url{http://dx.doi.org/10.1038/nphys4302}

\bibitem{Huang2022}
J.~Huang, R.~Fu, X.~Ye, D.~Fan, CCF Transactions on High Performance Computing
  \textbf{4}(1), 1 (2022).
\newblock \doi{10.1007/s42514-022-00089-w}.
\newblock \urlprefix\url{http://dx.doi.org/10.1007/s42514-022-00089-w}

\bibitem{Pagano2022}
S.~Pagano, C.~Barone, M.~Borghesi, W.~Chung, G.~Carapella, A.P. Caricato,
  I.~Carusotto, A.~Cian, D.D. Gioacchino, E.~Enrico, P.~Falferi, L.~Fasolo,
  M.~Faverzani, E.~Ferri, G.~Filatrella, C.~Gatti, A.~Giachero, D.~Giubertoni,
  A.~Greco, C.~Kutlu, A.~Leo, C.~Ligi, G.~Maccarrone, B.~Margesin, G.~Maruccio,
  A.~Matlashov, C.~Mauro, R.~Mezzena, A.G. Monteduro, A.~Nucciotti, L.~Oberto,
  V.~Pierro, L.~Piersanti, M.~Rajteri, A.~Rettaroli, S.~Rizzato, Y.K.
  Semertzidis, S.~Uchaikin, A.~Vinante, IEEE Transactions on Applied
  Superconductivity \textbf{32}(4), 1 (2022).
\newblock \doi{10.1109/tasc.2022.3145782}.
\newblock \urlprefix\url{http://dx.doi.org/10.1109/TASC.2022.3145782}

\bibitem{Sprengers2011}
J.P. Sprengers, A.~Gaggero, D.~Sahin, S.~Jahanmirinejad, G.~Frucci,
  F.~Mattioli, R.~Leoni, J.~Beetz, M.~Lermer, M.~Kamp, S.~H\"{o}fling,
  R.~Sanjines, A.~Fiore, Applied Physics Letters \textbf{99}(18) (2011).
\newblock \doi{10.1063/1.3657518}.
\newblock \urlprefix\url{http://dx.doi.org/10.1063/1.3657518}

\bibitem{Long2021}
K.R. Long, D.~Lucchesi, M.A. Palmer, N.~Pastrone, D.~Schulte, V.~Shiltsev,
  Nature Physics \textbf{17}(3), 289 (2021).
\newblock \doi{10.1038/s41567-020-01130-x}.
\newblock \urlprefix\url{http://dx.doi.org/10.1038/s41567-020-01130-x}

\bibitem{Gandolfi2012}
D.~Gandolfi, M.~Niedermayr, M.~Kumph, M.~Brownnutt, R.~Blatt, Review of
  Scientific Instruments \textbf{83}(8) (2012).
\newblock \doi{10.1063/1.4737889}.
\newblock \urlprefix\url{http://dx.doi.org/10.1063/1.4737889}

\end{thebibliography}

\end{document}